\documentclass[12pt]{iopart}
\usepackage{iopams}  
\usepackage{graphicx}
\usepackage{textcomp}

\graphicspath{{delay_graphs/}}

\begin{document}

\title[Phase locking to a LISA arm: first results on a hardware model]{\begin{center}
Phase locking to a LISA arm: \\
first results on a hardware model
\end{center}}

\author{Antonio F Garc\'ia Mar\'in$^1$, Gerhard Heinzel$^1$, Roland Schilling$^1$, Albrecht R\"udiger$^1$,
Vinzenz Wand$^1$, Frank Steier$^1$, Felipe Guzm\'an Cervantes$^1$, Andreas Weidner$^1$, Oliver Jennrich$^2$, Francisco J Meca Meca$^3$ and  
K. Danzmann$^{1,4}$} 
\address{$^1${} Max-Planck-Institut f\"ur Gravitationsphysik (Albert-Einstein-Institut),
Callinstrasse~38, D-30167 Hannover, Germany}
\address{$^2${} ESTEC, Noordwijk, The Netherlands}
\address{$^3${} Universidad de Alcal\'a de Henares Ctra. de Madrid-Barcelona, Km. 33,600
28871 Alcal\'a de Henares (Madrid), Spain}
\address{$^4${} also at: Universit\"at Hannover, Institut f\"ur Atom- und
Molek\"ulphysik, Callinstr.~38, D-30167 Hannover, Germany.}
\pacs{04.80.Nn, 95.55.Ym,  07.60.Ly, 07.87.+v, 42.30.Rx}
\begin{abstract}
We present the first experimental confirmation of the so-called ``self-phase-locked delay interferometry". This laser frequency stabilization technique consists basically in comparing the prompt laser signal with a delayed version of itself that has been reflected in another LISA satellite $5\cdot10^{9}\,\mathrm{m}$ away.
In our table-top experiment, the phase of a voltage controlled oscillator is stabilized by means of a control loop based on this technique. In agreement with the theory, the measured unity gain frequency is not limited by the inverse of the used delay (1.6\,$\mu$s).
In the time domain the system also behaves as predicted, including the appearance of a quasi-periodic ``ringing" just after the lock acquisition, which decays exponentially. Its initial amplitude is smaller when the loop gain is slowly ramped up instead of suddenly switched on.
\end{abstract}
\ead{antonio.garcia@aei.mpg.de}



\section{Motivation}

LISA is an ESA-NASA project to detect gravitational waves, involving three spacecraft flying in an equilateral triangle formation approximately 5 million kilometres apart. Together, they will act as a Michelson interferometer, covering a frequency range from 0.1\,mHz to 1\,Hz  and having a typical strain sensitivity of $10^{-23}$ \cite{weisses_buch}.

The phase noise of the LISA lasers would limit the sensitivity of the interferometer despite the use of traditional frequency stabilization techniques such as Pound-Drever-Hall with a stable reference cavity. TDI \cite{TDI} represents an option to overcome this problem by postprocessing the acquired data. Nevertheless, it requires high performance of the pre-stabilization methods mentioned before and it becomes more complicated when spacecraft motions are taken into account.

A traditional approach used with ground based detectors consists in locking the laser frequency to the arms of the interferometer. The LISA arms are good candidates for this technique due to their exceptional stability in the measurement frequency band, but the delay caused by the roundtrip travel time between two satellites (33\,s) had long been considered an insurmountable limitation. The control bandwith of this kind of loop is typically reduced to frequencies well below the inverse of this delay \cite{logan}, but that would mean a fraction of $30\,\rm mHz$ control bandwidth for LISA whereas very high gain at these frequencies is necessary to make the stabilization useful. Recently, some groups (see~\cite{Sheard2003},~\cite{Roland} and~\cite{Shoemaker}) have come up with control proposals and simulations achieving the necessary bandwidth and gain, what has been called the ``self-phase-locked delay interferometry". This paper describes an experimental demonstration of the principle of operation and the performance of the technique using an electrical model system. 

A voltage controlled oscillator (VCO) is stabilized in its frequency using a delay of $\tau = 1.6\,\mu$s realized by $300\,\rm m$ of coaxial cable, and exhibiting the highest unity gain frequency (UGF) of the control loop beyond $1/\tau$.

The predicted noise suppression was confirmed by a direct measurement of the oscillator's signal. In the time domain, a quasi-periodic, exponential decaying transient that was predicted to appear just after the lock acquisition (\cite{Sheard2003} and \cite{Tinto1}) could also be experimentally confirmed. Furthermore, its initial amplitude is reduced when the loop is closed by ramping up the gain instead of abruptly switching the loop on.

\section{Description and characterization of the system}

\begin{figure}[htb] \leavevmode
\begin{center}
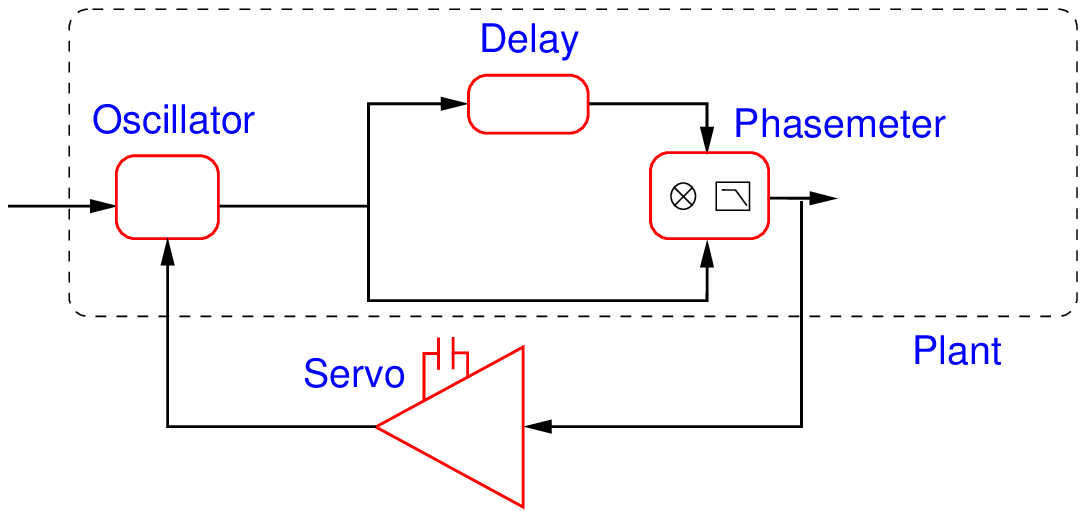
\caption{Principle of operation of the self-phase-locked delay interferometry.} \label{skizze1}
\end{center}
\end{figure}

Referring to Figure~\ref{skizze1}, an oscillator signal is split into two paths. After one of them has undergone a delay $\tau$, they are recombined, and a phasemeter detects their phase difference $\Delta\Phi$. When the loop is closed, $\Delta\Phi$ is used as the error signal $r(t)$ for a control system (servo) with frequency response $G(s)$. The output of the servo (feedback signal) $v(t)$ compensates the frequency fluctuations $p(t)$ of the oscillator . Finally, $q(t)$ is the frequency noise remaining in the stabilized system. In this paper the time will be called $t$ and we will use the Laplace variable $s=\rm i\omega$. Functions in the frequency domain are written with upper case letters and those in the time-domain in lower case letters. We will refer to the ``system" as the whole stabilization loop, consisting of the ``servo" and the ``plant".

In our experiment (see Figure~\ref{skizze2} and Table~\ref{table1}) the role of the LISA laser is played by a VCO working at approximately 72\,MHz. Instead of the two times 5$\cdot10^{9}$\,m pathlength between two LISA spacecraft, one of the signals goes through $300\, \rm m$ low-loss coaxial cable which causes a delay of $1.6\, \mu$s. The inverse of the delay is 625\,kHz, and the frequency range equivalent to the LISA measurement window goes from 2\,kHz to 20\,MHz.

\begin{table}
\caption{\label{table1}Correspondence between the LISA properties relevant for the experiment and our prototype}
\begin{indented}
\item[]\begin{tabular}{lll}
\br
 & LISA & Prototype\\
\mr
Signal & Laser & VCO\\
Delay $\tau$ & 33\,s & $1.6\,\mu$s\\
$1/\tau$ & 30\,mHz & 625\,kHz\\
freq.~range & 0.1\,mHz  1\,Hz & 2\,kHz  20\,MHz\\
\br
\end{tabular}
\end{indented}
\end{table}

In this section we will compare the theoretical response of the phasemeter $\Delta\Phi$ to frequency noise $p(t)$ of the oscillator with the one measured in our prototype in the open loop case. Note that we actuate on the frequency of the oscillator instead of its phase, which results in an extra factor of $1/f$ in the transfer function with respect to \cite{Sheard2003}. After that, we will discuss the open loop gain (OLG) and present the characteristics of the servo.

\subsection{Transfer function}
\label{sec_tf_teo}

The transfer function of the plant without servo (see Figure~\ref{skizze1}), measured from the frequency fluctuations of the oscillator $p(t)$ (expressed in $\rm rad/s$) to the phasemeter output~$\Delta\Phi$ (expressed in rad)  can be written as
\begin{equation}
H_{\rm theo}(\mathrm{i}\omega) = \frac{1-\mathrm{exp}(-\mathrm{i}\omega\tau)}{\mathrm{i}\omega} = \tau\:\frac{\mathrm{sin}(\omega\tau/2)}{(\omega\tau/2)}\:\mathrm{exp}(-\mathrm{i}\omega\tau/2).
\label{eq_tf_teo}
\end{equation} 

\begin{figure}[hbtp] \leavevmode
\includegraphics[width=16 cm]{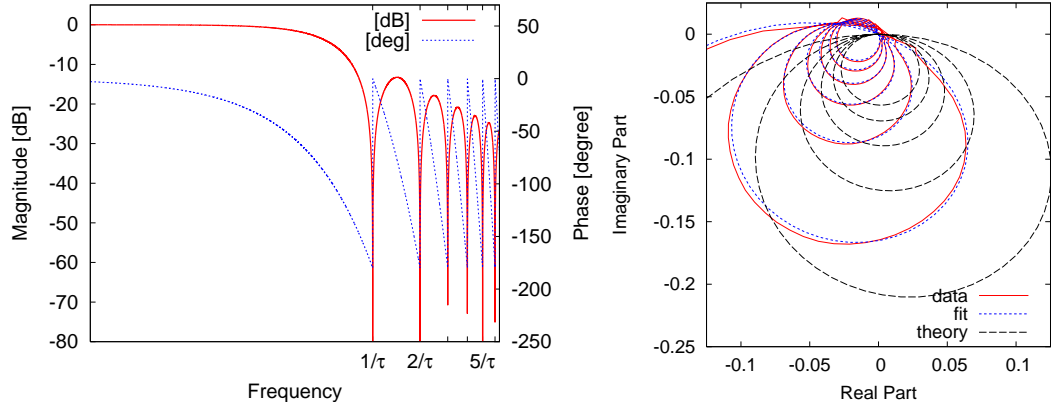}
\caption{Transfer function of the plant. Left: Bode representation of the theoretical transfer function. The magnitude is given in units of $\tau$. Right: Nyquist representation. The curve labeled ``theory" represents the theoretical transfer function. The curve labeled ``data" represents the one measured on the prototype and the curve labeled ``fit" shows the model presented in Equation~\ref{eq_tf_meas}.}
 \label{tf_teo}
\end{figure}

\begin{figure}[htbp]\leavevmode
\begin{center}
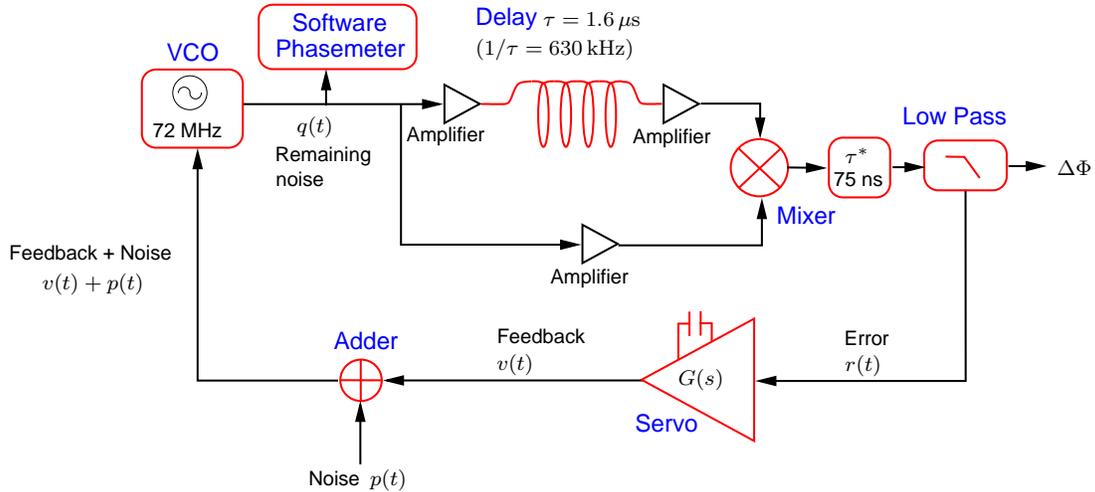
\caption{Table-top prototype: the phase measurement causes an extra delay $\tau^{*}$.}
\label{skizze2}
\end{center}
\end{figure}

Figure~\ref{tf_teo} shows the Nyquist and Bode representations of this function together with the data measured on the prototype. A model has been fitted to the data that additionally includes an extra delay $\tau^{*}$ of 75\,ns. This delay $\tau^{*}$ accounts for effects in the VCO, the short interferometer arm and the phasemeter. The total effect is modelled by one single delay $\tau^{*}$ at the phasemeter output (see Figure~\ref{skizze2}). The model also includes additional poles at $\omega_1, \omega_3$ and a zero at $\omega_2$ for the not ideal frequency response of the different components and is given by:

\begin{eqnarray}
\fl \eqalign{\ms H_{fit}(\mathrm{i}\omega) =\tau\frac{\mathrm{sin}(\omega\tau/2)}{(\omega\tau/2)}\mathrm{exp}(-\mathrm{i}\omega\tau/2) \mathrm{exp}(-\mathrm{i}\omega\tau^{*})
\left(\frac{1}{1+\frac{\rm{i}\omega}{\omega_{1}}}\right)
\left(1+\frac{\rm{i}\omega}{\omega_{2}}\right)
\left(\frac{1}{1+\frac{\rm{i}\omega}{\omega_{3}}}\right)\\
\mathrm{with}  \quad \omega_{1} = 2\pi\cdot530\,\mathrm{kHz} \quad \omega_{2} = 2\pi\cdot830\,\mathrm{kHz} \quad \omega_{3} = 2\pi\cdot12\,\mathrm{MHz} \quad \tau^{*} = 75\,\mathrm{ns}.}
\label{eq_tf_meas}
\end{eqnarray} 

Figure~\ref{skizze2} shows a more accurate description of the experimental setup including the extra delay $\tau^{*}$. The transfer function shown in Figure~\ref{tf_teo} was measured as 
\begin{equation}
H(\mathrm{i}\omega) = \frac{R(\mathrm{i}\omega)}{V(\mathrm{i}\omega)+ P(\mathrm{i}\omega)}
\end{equation}

\subsection{Open Loop Gain (OLG)}
\label{olg}
\begin{figure}[htbp] \leavevmode
\begin{center}
\includegraphics[width=8 cm]{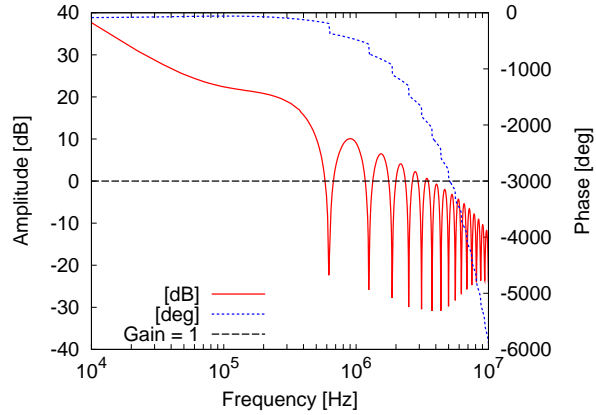}
\caption{Measured Open Loop Gain} 
\label{olg_meas}
\end{center}
\end{figure}

The Open Loop Gain of the system was measured in a stable, well-behaved loop as 
\begin{equation}
OLG(\mathrm{i}\omega)=H(\mathrm{i}\omega)G(\mathrm{i}\omega)= \frac{V(\mathrm{i}\omega)}{V(\mathrm{i}\omega)+P(\mathrm{i}\omega)}.
\end{equation} 
The highest unity gain frequency (UGF) takes place at about 3.5\,MHz (see Figure~\ref{olg_meas}), clearly above the inverse of the delay (625\,kHz). 

The noise suppression function predicted by such an OLG is given by:
\begin{equation}
C(\mathrm{i}\omega)=
\frac{1}{\left|1+OLG(\mathrm{i}\omega)\right|}=
\frac{1}{\left|1+G(\mathrm{i}\omega)H(\mathrm{i}\omega)\right|}.
\label{noise_sup}
\end{equation}

\begin{figure}[htbp] \leavevmode
\begin{center}
\includegraphics[width=18 cm]{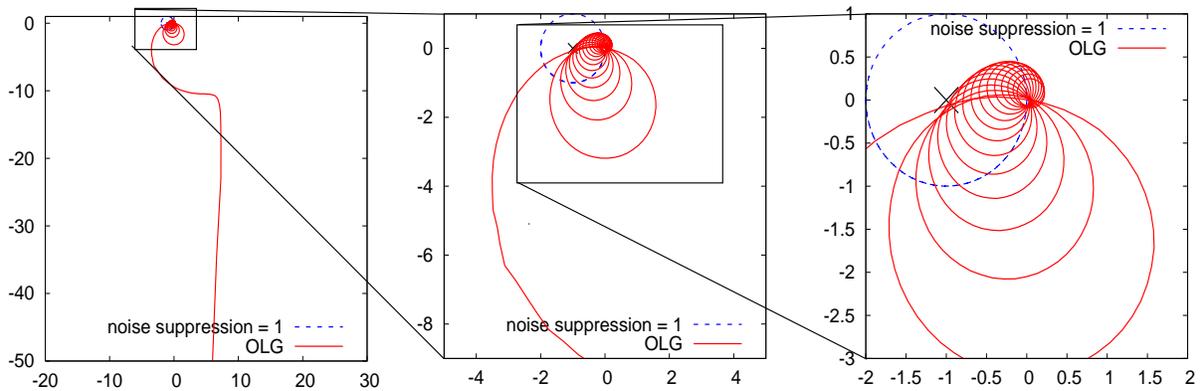}
\caption{Nyquist representation of the measured Open Loop Gain} 
\label{olg_nyquist}
\end{center}
\end{figure}

For a better interpretation we consider the Nyquist representation of the OLG shown in Figure~\ref{olg_nyquist}. The unity circle centred at ($-1$,0) represents the transition between noise suppression and noise enhancement. Besides, if the OLG encircles the ($-1$,0) point, the system becomes unstable. In this representation, the additional phase lag $\tau^{*}$ causes a clockwise rotation of the spiral-like gain curve and thus limits both the loop bandwidth and the gain of the servo. It is important to distinguish between this spurious delay $\tau^{*}$ that appears in the control loop after the signals have recombined and the intended delay $\tau$ that is applied to only one of the split signals.

\subsection{Controller}

\begin{figure}[htbp] \leavevmode
\begin{center}
\includegraphics[width=15 cm]{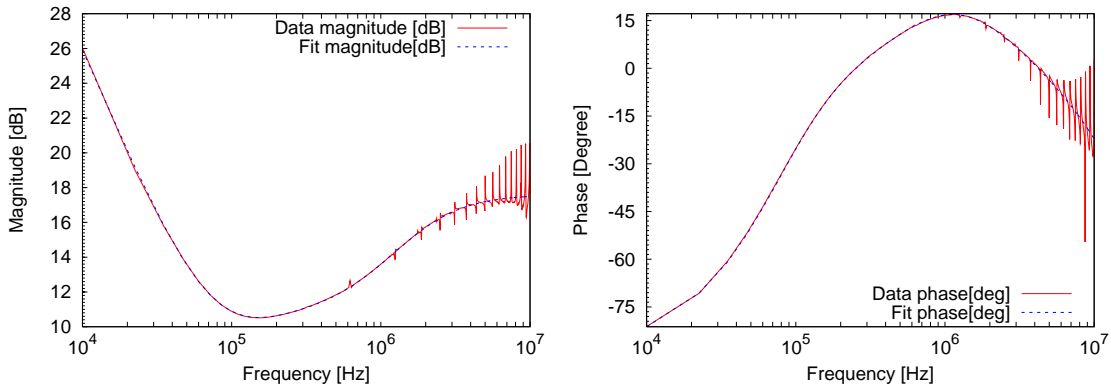}
\caption{Frequency response of the used servo and a fit to the measured data.} 
\label{servo_meas}
\end{center}
\end{figure}

Our controller (Figure~\ref{servo_meas}) consists of alternating poles and zeros at frequencies 106.6\,kHz, 172.4\,kHz, 843.7\,kHz, 1.8\,MHz, producing a frequency response approaching $f^{0.3}$ between 200\,kHz and 1\,MHz. Besides, there is an extra integrator from DC to 100\,kHz in order to increase the noise suppression in this region.

The starting point for such a controller design is the theory presented in \cite{Sheard2003}. The only difference is that Sheard et al.\ consider the frequency actuator as part of the controller rather than of the plant. This corresponds to taking a factor of $1/f$ from the transfer function presented here and giving it to the servo frequency response, which then becomes $f^{-0.7}$. 

\section{Results}
This section describes measurements of the noise behavior of the oscillator and compares them with the theory. The signal of the oscillator (72\,MHz) is directly sampled at 1\,GHz. A software phasemeter performs a Single Bin Discrete Fourier Transformation on it (\cite{ltp_paper},\cite{surrel}) and delivers a time series of the oscillator's phase $q(t)$ at 72\,MHz data rate (see Figure~\ref{skizze2}). We will study this signal in both the time and the frequency domain.

\subsection{Frequency domain}
\label{sec_noise}
\begin{center}
\begin{figure}[htbp] \leavevmode
\includegraphics[width=8 cm]{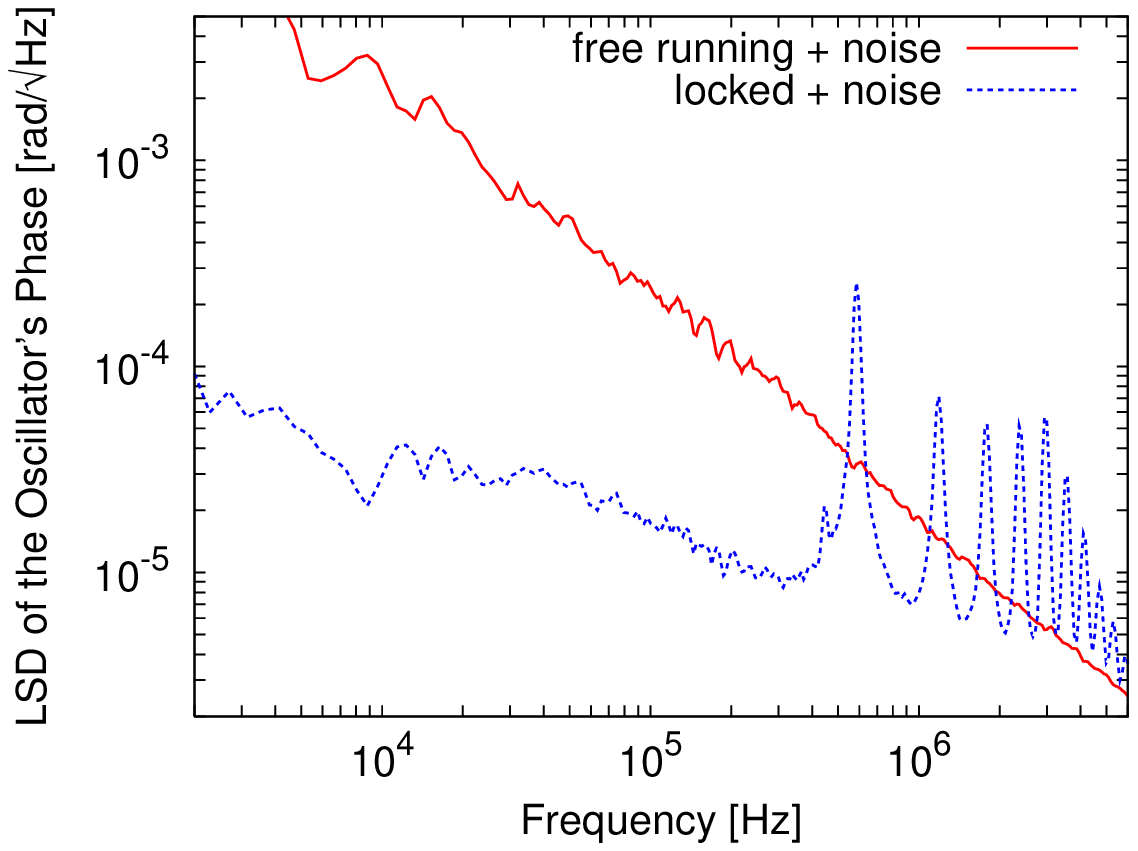}
\includegraphics[width=8 cm]{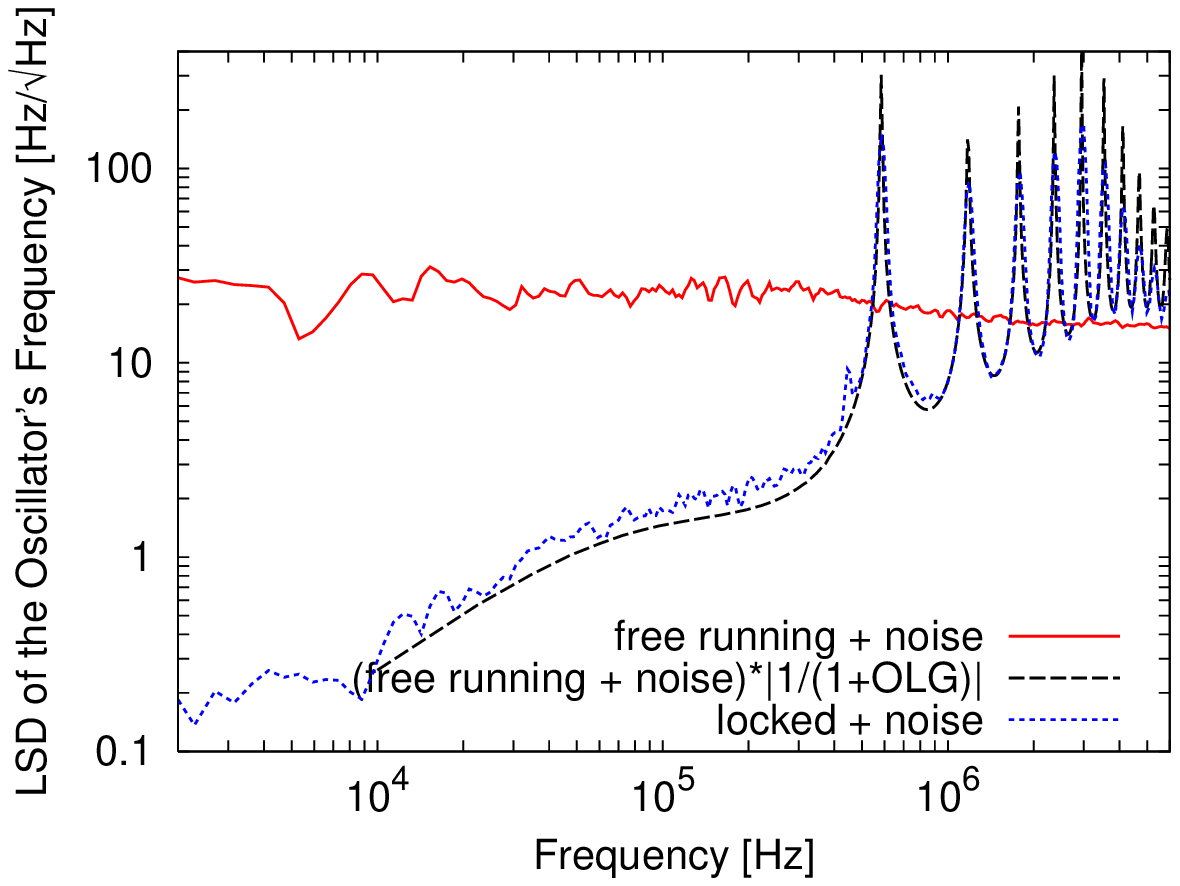}
\caption{Linear spectral density (LSD) of the oscillator's phase (left) and frequency (right). The solid curves show the oscillator in ``free-running" mode and the dotted ones refer to the stabilized state. The dashed curve on the right shows the frequency noise suppression predicted by the measured OLG (Section~\ref{olg}).} 
\label{noise}
\end{figure}
\end{center}
For the measurement shown in Figure~\ref{noise}, white frequency noise ($p(t)$ in Figure~\ref{skizze2}) has been added into the system thus generating phase noise with a $1/f$ linear spectral density (LSD). Once the system is stabilized, the remaining noise $Q(s)$ can be seen in the dotted curve which shows noise reduction at certain frequencies above 1/$\tau$. Superimposed to this curve, the disturbance sensitivity function 
\begin{center}
\begin{equation}
P(i\omega)C(i\omega)=\frac{P(i\omega)}{\left|1+OLG(i\omega)\right|}
\label{rem_noise}
\end{equation}
\end{center}
is plotted, where $P(i\omega)$ is the LSD of the introduced frequency noise and $OLG(i\omega)$ is the measured OLG presented in Section~\ref{olg}. The reasonably good agreement between the two curves confirms the predicted noise suppression.

\subsection{Time domain investigations}
The time evolution of the oscillator's phase during lock acquisition is analyzed here. For all the figures of this subsection, the controller is turned on at $t=0$ and time units are scaled to $\tau$. For the solid curves, the gain of the controller was turned on abruptly, whereas for the dashed ones the gain was ramped up linearly during approximately $16\,\mu \rm s\,(10\,\tau)$. 

\begin{figure}[htbp] \leavevmode
\begin{center}
\includegraphics[width=7 cm]{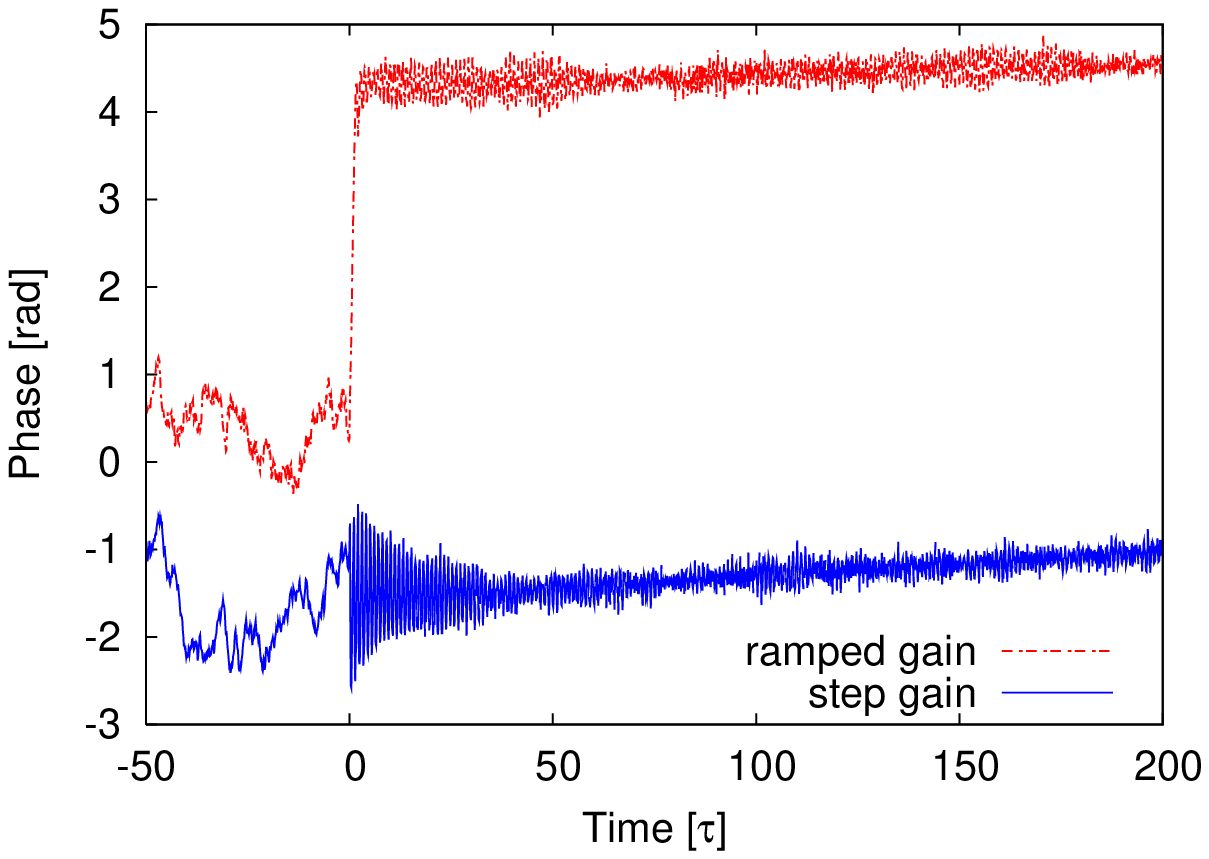}
\includegraphics[width=7 cm]{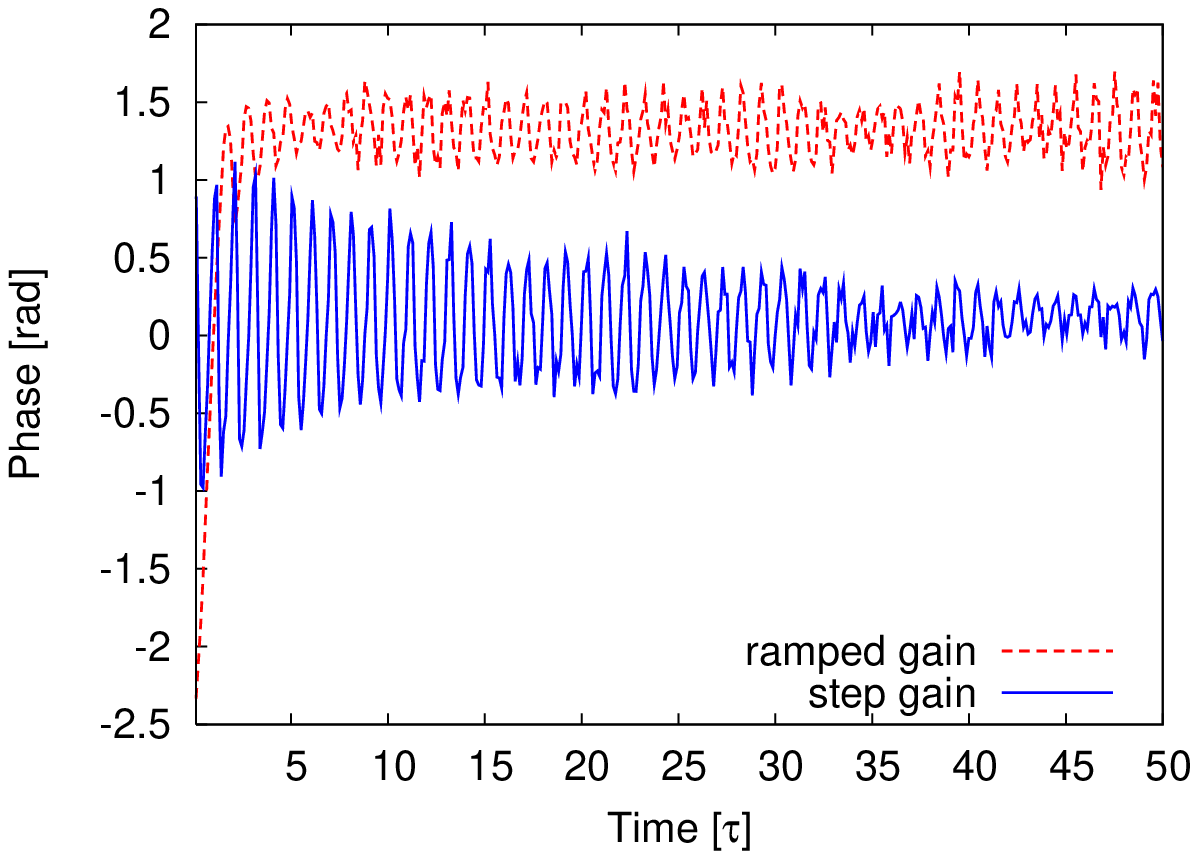}
\end{center}
\caption{Lock acquisition as white frequency noise is being added into the system. The initial amplitude of the transient is smaller when the gain is ramped up than when turned on abruptly. Left: General overview before and after the lock. Right: detailed view just after the lock. } 
\label{ramp}
\end{figure} 

Figure~\ref{ramp} shows the transient for these two cases with white frequency noise added as described in Section~\ref{sec_noise}. As predicted in~\cite{Sheard2003} and~\cite{Tinto1}, a pseudo-periodic transient can be observed just after the lock acquisition, whose initial amplitude is smaller when the gain is ramped up as opposed to the case of abrupt switching.

\begin{figure}[htbp] \leavevmode

\begin{center}
\includegraphics[width=7 cm]{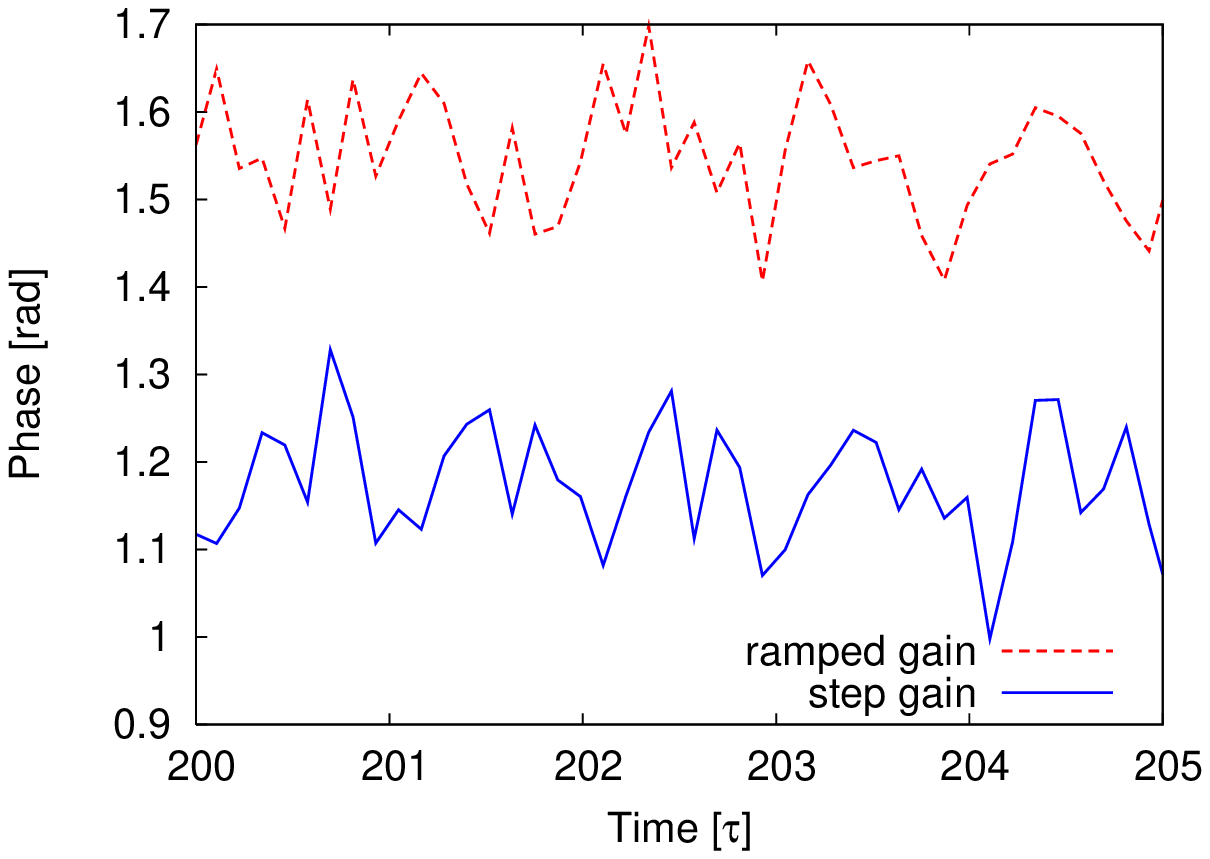}
\includegraphics[width=7 cm]{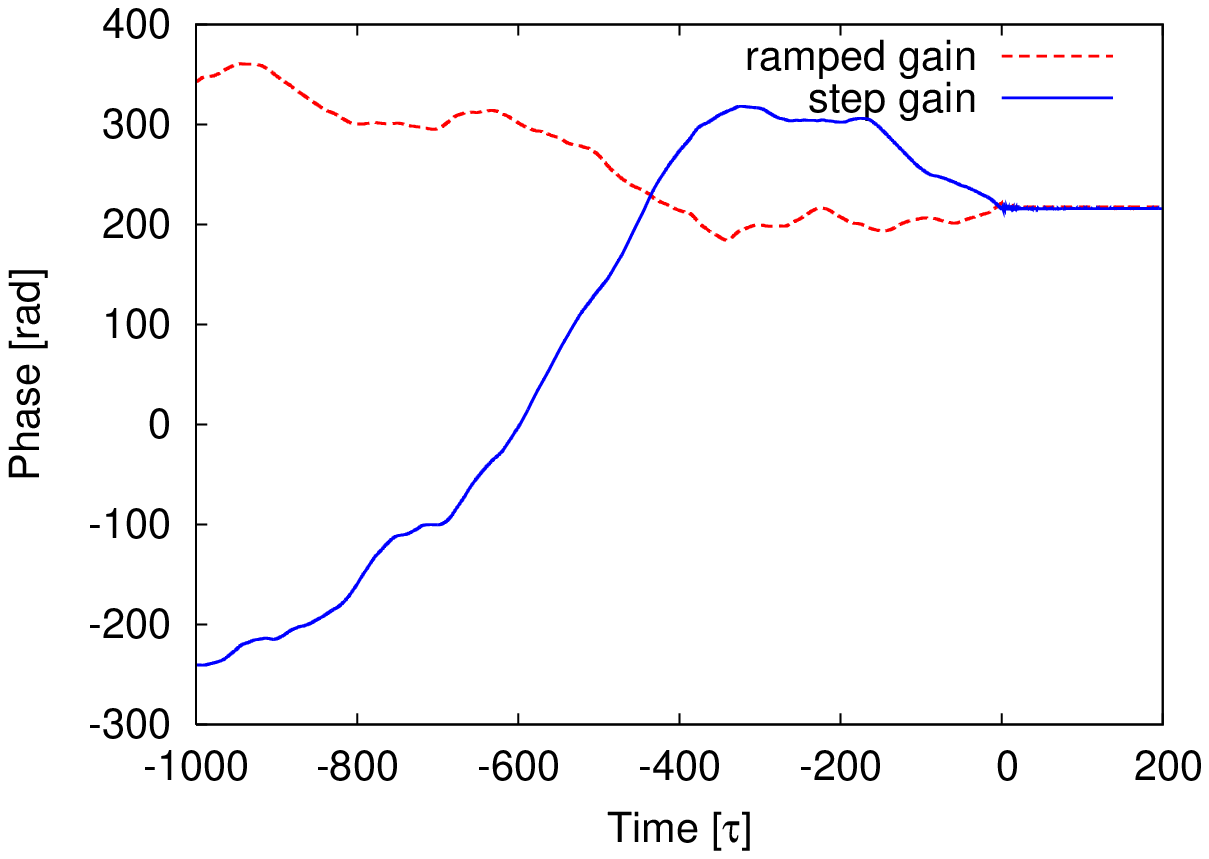}
\end{center}
\caption{Left: Detailed view of the time series shown in Figure~\ref{ramp}. It begins 200 $\tau$ after locking. Right: Lock acquisition in the presence of $1/f$ noise} 
\label{ramp2}
\end{figure}

The remaining noise is shown in the graph on the left in Figure~\ref{ramp2}. It does not show pure repetition but a structure typical to the sum of pseudo-harmonic narrow-band noise. This structure originates from the filtering done by the controller, as can be seen in Figure~\ref{noise}. 
On the right, $1/f$ noise is injected instead of white noise. It shows how the system locks despite the strong perturbations that drive the phase of the oscillator over several hundreds of radians before the stabilization is turned on.
\newpage 
\section{Discussion}

 
The ``self-phase-locked delay interferometry" (see~\cite{Sheard2003}) detects frequency fluctuations of a LISA laser by measuring the phase difference between the prompt laser signal and a delayed ($\tau = 33\,s$) version of it that has been reflected on a different LISA satellite. In the hardware model presented here, the phase substraction takes place between the signal from a VCO and a second version of it, delyed by $\tau = 1.6\,\mu$s. 
Frequency fluctuations of the VCO show up in our phase difference in the same way as frequency fluctuations of the laser do in the LISA configuration, which has allowed us the implementation of a frequency stabilization for the VCO based on the one described in~\cite{Sheard2003}. Although it takes place in a different frequency range due to the small delay of $1.6\,\mu$s, it permits the experimental confirmation of the main features of the ``self-phase-locked delay interferometry" .

First of all, the highest UGF of this kind of stabilizations is not limited to values far under 1/$\tau$, as was traditionally assumed (see~\cite{logan}). This can be seen from the measured OLG (Figure~\ref{olg_meas} and~\ref{olg_nyquist}) and from the noise plots (Figure~\ref{noise}), in which frequency noise reduction takes place at frequencies above 625\,kHz (1/$\tau$) and the highest UGF appears at about 3.5\,MHz.

Actually, the performance of our loop is only limited by the spurious delay $\tau^{*}=75$\,ns present at the output of the phasemeter. This delay $\tau^{*}$ limits the bandwith and gain of the servo as discussed in section~\ref{olg}. Such delays appear frequently in the experimental realization of a phasemeter and therefore care should be taken to minimize them in further implementations of this technique. 

The frequency noise of the oscillator gets reduced when the stabilization is turned on, as can be seen in Figure~\ref{noise} for the frequency domain and Figure~\ref{ramp} and~\ref{ramp2} for the time domain. This noise reduction is also in agreement with the performance that derives from the measured OLG (Figure~\ref{noise}).

Our hardware model of the``self-phase-locked delay interferometry" demonstrates that the 33\,s delay present in LISA does not represent a fundamental limitation in the performance of the stabilization. The combination of the technique discussed here, together with TDI and its recent improvements \cite{Shaddock}, may take us to a shot noise limited LISA without major hardware modifications in the actual baseline.

\end{document}